\documentclass[11pt]{article}

\usepackage{physics}
\usepackage{amsmath}
\usepackage{xcolor}
\usepackage{graphicx}
\usepackage{setspace}
\usepackage{placeins}
\usepackage{url}
\usepackage{authblk}

\title{Discrete Diffraction for Spectral Purification in Spontaneous Four Wave Mixing: a Recipe}

\author{Philip B. Main, Peter J. Mosley, Andrey V. Gorbach}
\affil{\it
 Centre for Photonics and Photonic Materials, 
 Department of Physics, University of Bath, Bath BA27AY, UK}

\newcommand{\fwm}{four wave mixing}

\begin{document}
\maketitle

\doublespacing

\begin{abstract}
Linear discrete diffraction of light within a waveguide array allows control over the longitudinal spatial profile of light. 
We propose a method of using this control to effectively modulate the nonlinearity in a spontaneous \fwm{} system. This allows the removal spurious frequency correlations between the generated signal and idler photon pairs.   
Weaving our method into a recipe for waveguide design, we demonstrate a spectral purity improvement from $\mathcal{P} =0.78$ to near saturation at $\mathcal{P} =0.97$, for a simple silicon-on-insulator geometry. 
\end{abstract}


Photon pairs generated by nonlinear processes are naturally accompanied by frequency entanglement. Such entanglement generates distinguishing information which unless properly managed, reduces their usefulness as single photon sources. 
This problem is intrinsic to the momentum conservation conditions within the nonlinear device, manifest in correlations within the phasematching envelope function $\Phi(\omega_s, \omega_i)$ of the photon-pair state vector \cite{Grice1997}: 
\begin{equation}
\ket{\psi_2} = \int d\omega_s d\omega_i \: f(\omega_s, \omega_i) \ket{\omega_s} \ket{\omega_i},
\end{equation}
where $\omega_s$ and $\omega_i$ denote the signal and idler frequencies. The frequency of the signal and idler photons are naturally anti-correlated by the frequency matching conditions of the process. For a \fwm{} source pumped at frequencies $\omega_m$ and $\omega_a$ this condition is given:    
\begin{equation}
0 = \omega_{m} + \omega_{a} -\omega_s - \omega_i
\label{eq:freq}
\end{equation}  
where $\omega_{m}$ and $\omega_{a}$ are the frequencies of the main and auxillary pumps respectively. 
 
However, a source capable of heralding indistinguishable photons is one in which the state vector can be expressed in a factored form \cite{Law2000, Grice2001, braczyk2017}: 
$ \ket{\psi} = \int d\Omega_1  \ket{u(\Omega_1)} \int d\Omega_2 \ket{v(\Omega_2)}$, which is prohibited by the solution of the phasematching function for a waveguide with step-like nonlinearity of strength $\gamma_0$: 
\begin{equation}
\Phi(\omega_s, \omega_i)  \propto \int_0^L dz \: \gamma(z) e^{i\Delta \beta(\omega_s, \omega_i) z} = i \gamma_0 L sinc[\frac{\Delta \beta(\omega_s, \omega_i) L}{2}],
\end{equation}
where the photons' bandwidths $\frac{\Delta\beta L}{2}$, set by the phase mismatch and length of the device, is reasonably broad. This is a result of the secondary maxima of the sinc function creating additional frequency correlated regions. 

This limitation has led to the development of many nonlinear apodisation techniques. These aim to produce uncorrelated photons using a nonlinear profile $\gamma(z)$ which transforms to a function without any secondary maxima. Such techniques particularly suit spontaneous down conversion sources, since the strong $\chi_{2}$ nonlinearity on which these sources rely, only exists (as a bulk property) in inversion anti-symmetric materials and can be removed or reversed by changing the crystal structure with minimal impact on the linear susceptibility. Thus the intrinsic link between the crystal structure and $\chi_{2}$, allows longitudinal control of the nonlinearity to be introduced by domain engineering of the material \cite{Branczyk2011, BenDixon2013, Dosseva2016, Chen2017, Graffitti2017}. 

This principle cannot be easily transferred to spontaneous \fwm{} sources since in this no such intrinsic link between material symmetry and nonlinearity exists. Therefore in situations where this type of source is advantageous, such as in all-fibre single photon sources \cite{McMillan2009, Francis-Jones2016} or an integrated quantum photonics platform \cite{Politi2009, Bogdanov2017}, this must be compensated for using significantly longer interaction lengths, spectral filtering, or pulse walk-off\cite{Fang2013}. The last of these methods takes advantage of the dynamics of a pair of pulsed bright light pumps to modulate the temporal overlap and remove correlations from the joint spectrum.  

We propose a method in which the spontaneous \fwm{} interaction is controlled by the spatial dynamics of the second cw pump in a waveguide array rather than the temporal dynamics of a pulsed pump. As well as not requiring the additional experimental complications accompanying the use of two pulsed pumps our method is advantageous for on-chip setups where waveguide integration is necessary. Similar methods have previously been proposed to modify the phasematching properties of nonlinear waveguides \cite{Dromey2007, Zhong2016, Setzpfandt2016, Saleh2018}. However, these operate in a different regime which requires much faster coupling between waveguides $L_c \approx \mathcal{O}(\frac{2\pi}{\beta})$, in contrast to our proposal where the coupling is of the order of the pair generation length $L_c < \mathcal{O}(P_z L^2)$.     



Our scheme relies on two things: a main pulsed pump which is confined to a single waveguide, and a second auxiliary cw pump which is free to couple across a waveguide array structure. 
Excitations of the auxiliary and main pumps in different waveguides means that the power of the auxillary pump will spread to neighbouring waveguides while the pump remains locally imprisoned.
As the two pumps begin to overlap spatially with the main pump, spontaneous \fwm{} can occur, producing pairs of photons. Since the amplitude profile of the auxiliary pump is changing due to diffraction within the waveguide array, the strength of the local nonlinear interaction is modulated along the length of the device. Therefore the dynamics of the auxiliary pump and the way it walks across the main pump can be used to apodise the phasematching function.

To model this local spontaneous \fwm{} we postulate the following Hamiltonian assuming the frequency matching condition from equation \ref{eq:freq} with a z dependant auxiliary pump amplitude $A_a(z)$:
\begin{equation}
\mathcal{H}_{nl}(z) =  \iint d\omega_s d\omega_i  
\: \gamma_{fwm} A_m A_a(z) \exp [i\Delta \beta z] \: \hat{a}_s^\dagger (\omega_s) \hat{a}_i^\dagger (\omega_i)+ H.c,
\end{equation}
where the nonlinear phasematching condition is given by $\Delta\beta = \beta_m + \beta_a - \beta_s - \beta_i$, the z profile of the auxillary pump $\alpha_a(z)$, the signal and idler frequency creation operators $\hat{a}_s^\dagger \hat{a}_s^\dagger$, the interaction strength $\gamma_{fwm}$. This Hamiltonian describes the creation of a pair of photons from two bright light fields $\alpha_m$ and $\alpha_a$ under the undepleted pump assumption. Here the canonical coordinate z is defined by the frame transformation $0 = z -  v_g t$ \cite{luks2009, mandel1995} ($v_g$ being the group velocity of light propagating in the fibre).

\begin{figure}[t!]
\centering
\includegraphics[scale = 1]{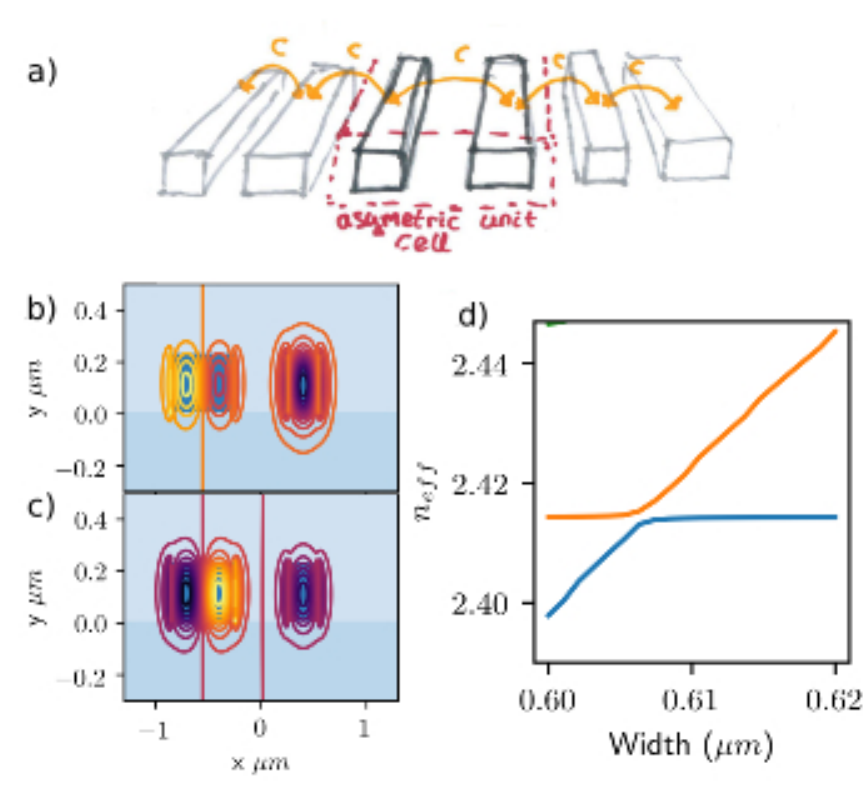}
\caption{
Construction of waveguide array to produce narrow bandwidth discrete diffraction. 
a) Schematic of array. Constructed by periodically repeating a unit cell designed to confine intra-waveguide coupling to a narrow band of wavelengths, our array consists of alternating waveguides of constant widths and separations.
b) and c) symmetric and antisymmetric supermodes (respectively) forming at a wavelength $\lambda_a = 1.3 \mu m$ and widths of $0.301\mu m$ $0.650\mu m$ calculated using our own finite difference frequency domain solver \cite{Main2017}.
d) Dispersion of supermodes as the width of the broader waveguide is varied for a fixed wavelength $\lambda_a = 1.3 \mu m$.
\label{fig:unit_cell}}
\end{figure} 
To realise the z dependant the auxiliary pump amplitude profile $A_a(z)$ we introduce an array of waveguides designed such that coupling is only phasematched for a narrow band of wavelengths around the auxiliary pump. This arises naturally within a waveguide array constructed from waveguides of alternating widths (see Fig.~\ref{fig:unit_cell}a). The key properties of this array: linear phasematching and coupling strength can be calculated from the unit cell which consists of an asymmetric coupler (as shown in Fig. \ref{fig:unit_cell}a). 
This asymetric coupler formed from two waveguides of different widths $w_m$ and $w_a$ (for main and auxillary respectively), enables hybridisation between modes of different order. In our case this is the TE10 mode of the narrow waveguide and the TE20 mode of the broad waveguide. 
Since these modes cross at a single point they will couple  only over a narrow range of widths and wavelengths \cite{snyder2012, Main2019}.
At the point of degeneracy the coupling coefficient is given by the difference between the supermodes:
\begin{equation}
C = \frac{\beta_{even} - \beta_{odd}}{2} \label{eq:couple}
\end{equation}
As can seen from fig. \ref{fig:unit_cell}d this point can be specified for a specific wavelength by sweeping the width of the auxiliary waveguide.
When tiled into a waveguide array the auxiliary pump will spread between waveguides with a longitudinally varying amplitude envelope.
The discrete diffraction (so named for analogues between the light behaviour and  electrons in the tight binding model \cite{Jones1965, Jena2003a},) dynamics of light in waveguide arrays has already been systematically studied in both the linear and nonlinear regimes allowing for many different possible spacial configurations of the auxiliary pump.

In the scope of this work we will focus on the simplest case: n equally spaced waveguides. The evolution of this system is governed by the nearest neighbour coupling matrix
\cite{Jena2003a}:
\begin{equation}
-i\frac{d}{dz}
\begin{bmatrix}
A_0(z) \\
\vdots \\
\vdots \\
A_n(z)
\end{bmatrix}
 = 
\begin{bmatrix}
& & C &  \\
C & & & \ddots &  \\
&\ddots&  & &  C& \\
& & C & & \\
\end{bmatrix}
\begin{bmatrix}
A_0(z) \\
\vdots \\
\vdots \\
A_n(z)
\end{bmatrix}. \label{eq:couple_waveguides}
\end{equation}
Here the waveguide coupling as defined in Eqn.~\ref{eq:couple}, remains constant across the structure and each waveguide  couples only to its nearest neighbours. The general solution for the z dependent amplitude envelope may be written: 
\begin{equation}
{\bf A}_a(z) = \sum c_i  e^{i \lambda_i z} {\bf \nu}_i
\end{equation}
where ${\bf \nu}_i$ and $\lambda_i$ are the eigenvalues and vectors (stationary solutions) respectively, of the coupling matrix and the coefficients $c_i$ are defined based on the boundary excitations of the structure.  

\begin{figure}[t]
\centering
\includegraphics[scale = .6]{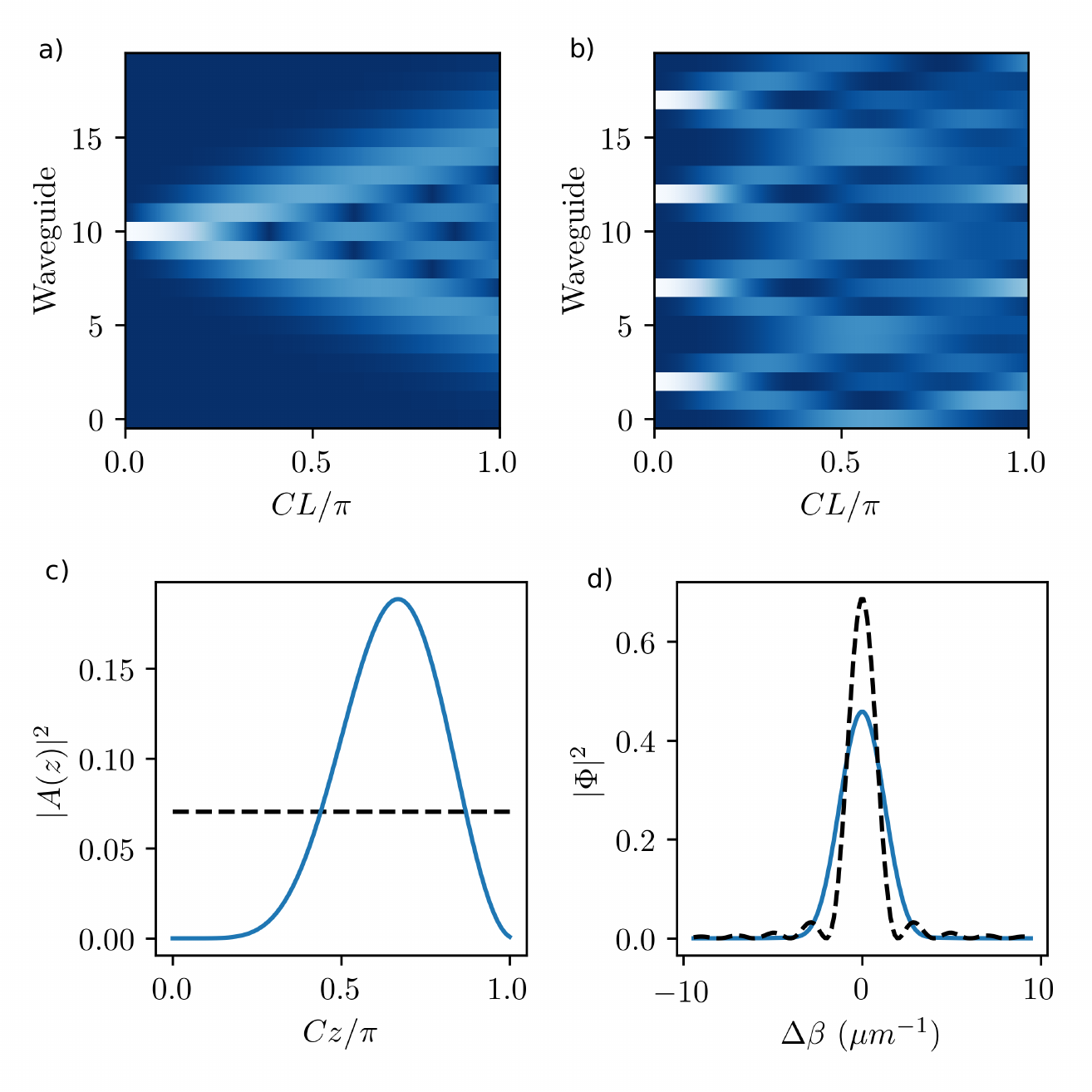}
\caption{Discrete diffraction and joint spectral control: a and b demonstrate distribution of power in a 20 waveguide array for single and periodic excitation respectively. c) shows the power profile of waveguide no.~7 from the single excitation with a normalised constant amplitude profile plotted with solid and dashed lines respectively. The joint spectral intensity $|\Phi|^2$ for each of these amplitude profiles is plotted in d).\label{fig:waveguide_diffract}}
\end{figure}
Fig.~\ref{fig:waveguide_diffract}a and b  plot the solution for two different boundary excitations of such a structure consisting of 20 waveguides: a single excitation and a periodic excitation. Using the amplitude profile of waveguide No. 7 it can be seen in Fig. \ref{fig:waveguide_diffract}d) that that the Joint spectral intensity $\Phi$ for generated photon pairs is strongly modified and the secondary sidelobes are suppressed. By combining the previous threads into a series of steps we will show how they can be integrated into a real world waveguide system. 

Our method can be summarised with a three-step recipe. ( Note that these results are designed as proof of principle rather than describing the definitive geometry for a practical system.) 

\begin{enumerate}
    \item {
Choose the geometry of the primary waveguide such that favourable phasematching contours exist. 

Silicon-on-insulator waveguides are strongly subwavelength. Correct balancing of geometric parameters (such as the aspect ratio \cite{Turner2006,Turner-Foster2010}) can be used to control the dispersive properties of the waveguide and arrange the phasematching between pump signal and idler. We select a simple SOI waveguide with a height of $0.22\mu m$ and a width of $0.3\mu m$ and predict that such a structure satisfies the group velocity constraints to allow for perpendicular intersection between the phase and frequency matching conditions on the signal-idler frequency plane. This group velocity matching constraint can be written \cite{Garay-Palmett2008}:
\begin{equation}
\frac{2}{v_m} -\frac{1}{v_s}-\frac{1}{v_i} = 0,  
\end{equation}
and we predict it is satisfied for combination of: auxiliary pump at $1.37 \mu m$, main pump centred at $1.17\mu m$ phasematching photon pairs at $\lambda_s = 1.54\mu m$, $\lambda_i = 1.08\mu m$ as shown in Fig.\ref{fig:phasematch_coupler}. 

\begin{figure}[tb]
\centering
\includegraphics[scale = .6]{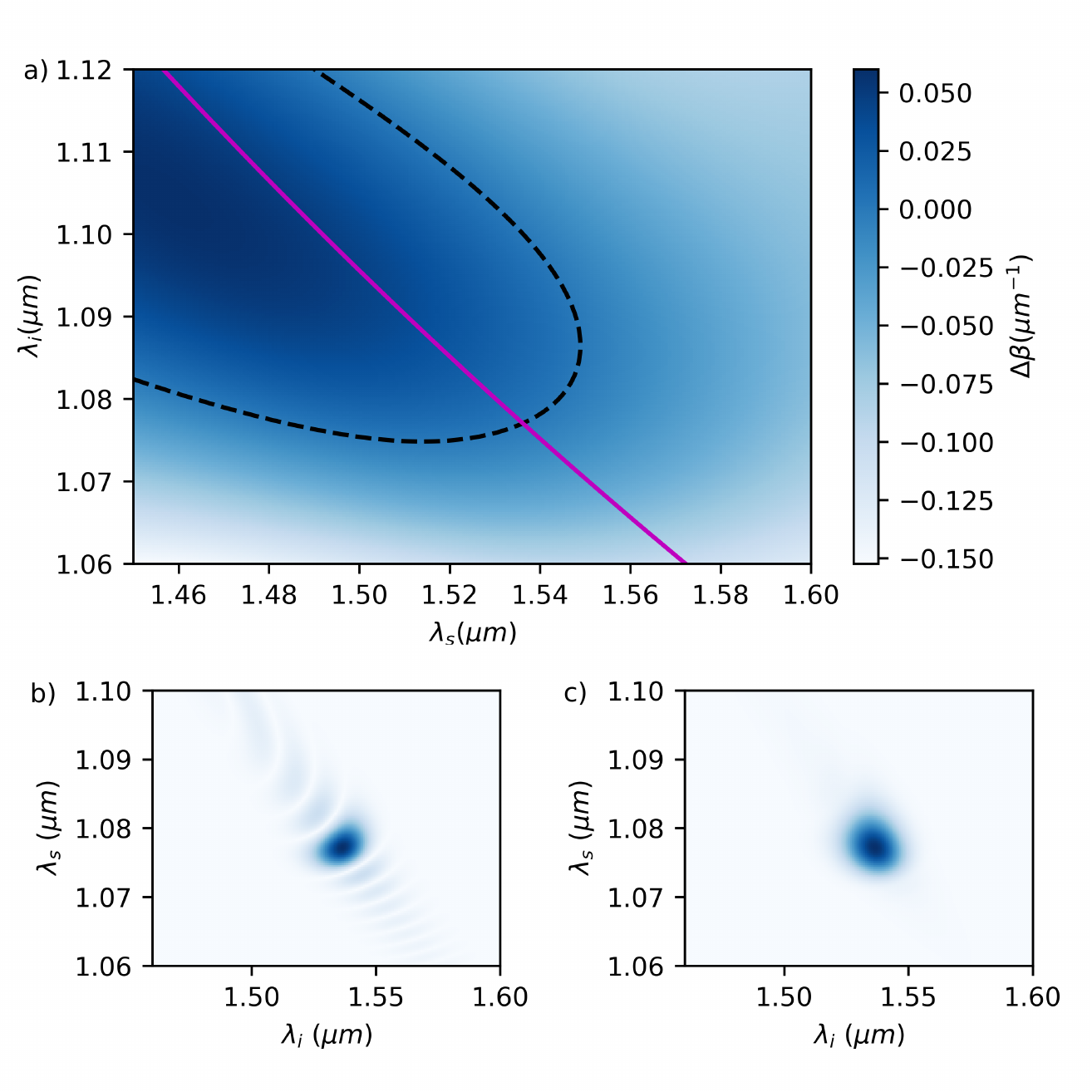}
\caption{Phasematching contour for an uncoupled $0.3\mu m$ width silicon on insulator waveguide and an auxillary pump fixed to $\lambda_a = 1.37\mu m$ . Here the frequency matching condition for the optimal main pump ($\lambda_m = 1.17\mu m$) is indicated by the pink contour. \label{fig:phasematch_coupler}}
\end{figure}
}

    \item {
Choose the auxiliary waveguide. 

As the second waveguide in the unit cell, this waveguide determines the pass wavelength and must be tuned such that coupling can occur around the auxiliary pump wavelength which gives the best phasematching. To achieve this we set the width of this waveguide such that coupling is phasematched between a higher order x-polarised mode and the fundamental mode of the primary waveguide. Fig. \ref{fig:unit_cell}d shows the anti crossing of these modes as the width of the secondary waveguide is varied. Selecting a width of $0.650\mu m$ the modes are phasematched and coupling can occur across the array. (See also Fig.\ref{fig:unit_cell}b and c which show the resulting supermodes for this width.)    
}


    \item { 
Match the coupling strength by selecting the inter-waveguide separation. 

The strength of the coupling is independently determined by the separation between waveguides and can be used to choose the desired amplitude profile of the diffracting pump.
For example, to achieve a coupling length, $L_c = \frac{\pi}{C}$ , of $500\mu m$ based on our geometry we require the waveguides to be separated by approximately $0.4\mu m$; or for $1000 \mu m$ a separation of approx $0.47 \mu m$. There is some flexibility here since there are two independent variables, L and  C to tune. However the optimal bandwidth of the main pump will be determined by the total length of the device.  
}

\end{enumerate}

Based on the geometry outlined in the previous step the degree of spectral entanglement or purity between the signal and idler photons can be calculated for a given main pump bandwidth. This is accomplished by the Schmidt decomposition of the joint spectra plotted in in Fig. \ref{fig:phasematch_coupler} defined: 
\begin{equation}
\mathcal{P} = tr[(\ket{\psi}\bra{\psi})^2] = \sum_k q_k^4, 
\end{equation} 
where $q_k$ is the Schmidt coefficient resulting from the decomposition. 
The joint spectra plotted in Fig.\ref{fig:phasematch_coupler}(b and c) have purities of $\mathcal{P} = 0.78$ and $\mathcal{P} = 0.97$ respectively. Thus we can see that the effective modulation of the nonlinearity generated by the diffraction of the secondary pump gives a significant improvement to the purity of the photons generated within rectangular silicon-on-insulator waveguides.   

In summary, we have proposed a new scheme for generating high purity single photons via \fwm{} in a waveguide.
Our proposed method allows effective nonlinear modulation; overcoming one of the roadblocks for very high purity heralded single photon generation in an integrated $\chi_3$ system.
In a practical waveguide array system we have shown how modulation by the discrete diffraction of an auxiliary pump, produces a significant improvement to the spectral purity over the baseline waveguide (with a steplike nonlinear profile).
  
This scheme has the advantages of being extensible and reconfigurable; the purity of photons is controlled by the diffraction of one or multiple pumps through a waveguide. 
We anticipate future directions of this work to include exploration of reconfigurability made possible by exciting different combinations of waveguides with the auxiliary pump.  
In addition, the relaxation of the uniform coupling assumption would introduce more control over the longitudinal profile of the auxiliary pump and provide an additional degree of freedom to fine tune the system. 

The device we propose could also naturally integrate into a spatial multiplexing scheme by pumping at multiple points in the array to produce a repeating diffraction pattern. This would allow a battery of heralding sources to be created by introducing the pulsed pump to the main waveguide at periodic intervals.

\bibliographystyle{ieeetr}
\bibliography{knowledge}

\end{document}